\documentclass[12pt]{article}
\usepackage{amsmath,amsthm,amssymb,epsfig,verbatim,hpkfinal}
\input a30.defs
\begin{document}
\title{Cosmological Einstein-Yang-Mills equations
\thanks{Supported in part by NSERC grant A8059.}
}
\author{
B.K. Darian\thanks{darian@phys.ualberta.ca}\\
Theoretical Physics Institute, University of Alberta\\
Edmonton, Canada T6G 2J1\\[2pt]
\and \\[2pt]
H.P. K\"{u}nzle\thanks{hp.kunzle.ualberta.ca}\\
Department of Mathematical Sciences, University of Alberta\\
Edmonton, Canada T6G 2G1}
\date{}

\newcommand{\sauthor}{Darian and K\"unzle}
\newcommand{\stitle}{Cosmological Einstein-Yang-Mills equations}

\maketitle

\begin{abstract}

We use a systematic construction method for invariant connections on
homogeneous spaces to find the Einstein-$SU(n)$-Yang-Mills equations
for Friedmann-Robertson-Walker and locally rotationally symmetric
homogeneous cosmologies.  These connections depend on the choice of a
homomorphism from the isotropy group into the gauge group. We consider
here the cases of the gauge group $SU(n)$ and $SO(n)$ where these
homomorphisms correspond to unitary or orthogonal representations of
the isotropy group. For some of the simpler cases the full system of
the evolution equations are derived, for others we only determine the
number of dynamical variables that remain after some mild fixing of
the gauge.

\end{abstract}

\sect{intro}{Introduction}

There has been extensive work on Einstein-Yang-Mills (EYM)
cosmological models in the last decade.  This work was partly
\mnote{I like 'was' better here}
motivated by the successes of inflationary models driven by scalar
fields in solving flatness and (to large extent) horizon problems in
cosmology.  The interest in inflationary models driven by
fields other than scalar fields is a consequence of less attractive
features of the former \cite{k5872}.
Mini-superspace EYM cosmology is a natural extension of a
non-perturbative treatment of self-gravitating scalar fields.  It has been
 realized that despite a large phase space
associated with seemingly redundant extra gauge degrees of freedom,
there already exists a systematic mathematical method (based on
Wang's theorem \cite{k4872}) for the construction of invariant
connections over homogeneous spaces in the same spirit as that of
Kaluza-Klein theories.  As will be seen in section~\ref{EYMhst}, such
invariant connections are related to the representation theory of
compact Lie algebras.  For some of the most easily constructed cases
of $SO(n)$-YM fields the solutions were obtained on closed
\mnote{Here I don't care too much, but I am not sure 'has been' is better}
Friedman-Robertson-Walker (FRW) cosmologies \cite{k1529},\cite{k5405},\cite
{k5996}.
$SU(2)$-YM fields on open FRW cosmologies have also
been of some interest \cite{k5209},\cite{k5873}. In this particular
representation there is one degree of freedom associated with the YM fields.

Conformal invariance of YM field equations (due to the fact that they
are zero-rest-mass fields) results for the homogeneous and isotropic
case in a decoupling of the gravitational and YM degrees of freedom.
The energy momentum tensor is that of a
radiation perfect fluid and the geometry is that of a Tolman universe.

Despite the fact that it is known that the construction of invariant YM
connections could be generalized --- at least in principle --- to other
compact gauge groups and cosmological models with compact and
non-compact spatial sections, a
systematic attempt to study models based on more complicated
representations in FRW and anisotropic homogeneous cosmologies has not
been conducted.

In
the present article we derive the EYM equations for $SU(n)$-FRW and
$SU(n)$ locally rotationally symmetric (LRS) cosmologies.

Section~\ref{EYMhst} is an exposition
of a general but rather explicit construction of the
Riemann and YM curvatures based on the theory of connections invariant
under symmetry groups that act transitively on the base manifold.  It
turns out that the resulting purely algebraic Yang-Mills equations do 
\mnote{I think we should avoid using `of course', `here', `now' as much as 
possible. I agree with regard to 'of course', but 'here' and 'now' are often
needed for a smooth flow of the text.}
not require any 
explicit choice of gauge. Such
space-time homogeneous models are not considered to be
realistic physically and we make no attempt in this paper to find any
exact solutions.

In section~\ref{EYMshst} we derive the EYM equations for spatially
homogeneous cosmological models. The result is a system of ordinary
differential equations where again the YM gauge needs to be fixed only
mildly, for example, by setting the temporal component of the
potential to zero. The spatially homogeneous and isotropic models are
discussed in section~\ref{isotropic}. Although the space-time geometry
is completely determined independently of the YM fields, the latter
satisfy
in general some complicated coupled system of evolution equations. We
derive here a few general facts for arbitrary gauge groups and some
more explicit equations corresponding to different possible YM fields
for the gauge groups $SU(n)$ and $SO(n)$.

Finally, in section~\ref{lrs} we consider, in a unified way, all
LRS cosmological models with a $SU(n)$ Yang-Mills source.
In such models, after solving for the constraints, there are $2(n-1)$ degrees
of freedom associated with the YM fields.
Here we just concentrate on what we consider the simplest
YM-connections that contain a 'magnetic' part and derive the full
evolution equations of the EYM-system. An analysis of the solutions of
this quite complicated system is beyond the scope of this paper. Even
the system for
homogeneous YM fields in two-dimensional flat space is known to be non-integrable.  A
dynamical system analysis of LRS Bianchi~I models with $SU(2)$-YM
fields was given in Ref. \cite{hka29}.

\sect{EYMhst}{Einstein-Yang-Mills equations on homogeneous space-time}

Following the conventions of
Ref. \cite{k0930} we let $(M,g)$ be a connected pseudo-Riemannian manifold
with its Levi-Civita connection, and $K$ its isometry group. Its
(left) action,
\leqn{act}{
\psb: K\times M \ra M: (a,x) \mapsto \psb_a x
}
is transitive and effective on $M$.  Fixing a
point $x_0\in M$ (to be called the {\em origin}) the isotropy subgroup
$K_0$ of $K$ is defined by
\leqn{isosg}{
K_0 := \{ a\in K \,|\, \psb_a x_0 = x_0\}
}
and all isotropy subgroups
for different points of $M$ are conjugate. The manifold $M$ is
diffeomorphic to the set of left cosets of $K$ with respect to $K_0$,
$M\cong K/K_0$, and there is a one-to-one correspondence between
$K$-invariant pseudo-Riemannian metrics on $M$ and
$\ad_{K_0}$-invariant non-degenerate symmetric bilinear forms $\go$ on
the quotient space $\kf/\kf_0$ of the corresponding Lie algebras.

We wish to describe Yang-Mills connections that have as many
symmetries as the metric of space-time and therefore assume that the
full isometry group also acts by principal bundle automorphisms
\leqn{PBauto}{
        \pst : K \times P \ra P
}
on the principal bundle $P$ that project onto isometries on $M$ thus satisfying
\leqn{PBauto1}{
        \pi\circ\pst = \psb\circ\pi \quad\text{and}\quad  \pst_a\circ R_g = R_g\circ\pst_a \quad\forall\; a\in K\,\,\forall\; g\in G
}
where $\pi$ is the projection, $G$ the structure (gauge) group, and
$R$ the right action of $P$. If the gauge potential is invariant under
this action, i.e. if the connection form $\omt$ on $P$ is invariant,
$\pst_a^*\omt = \omt$ for all $a\in K$, then so is the curvature
form $\Omt$, $\pst_a^*\Omt = \Omt$. It follows that
\leqn{ominv}{
  \Lie_{\Xt}\omt = 0 \quad\text{and}\quad  \Lie_{\Xt}\Omt = 0 \quad \forall\; X\in\kf.
}
where $\Xt$ is the infinitesimal generator of the action $\pst$ on $P$ corresponding to $X\in\kf$.

Now it is known (see, for example, Ref. \cite{k0832}) that equivalence
classes of such $\pst$-invariant principal bundles $P$ over $M$ are in
one-to-one correspondence with conjugacy classes of homomorphisms
$\lambda:K_0\ra G$. Here $\lambda$ and $\pst$ are related by
\leqn{lampsi}{
\pst_a(u_0) = R_{\lambda(a)}u_0 \quad \forall\; a\in K_0
}
where $u_0$ is any fixed element of $\pi^{-1}(x_0)$.

Moreover, Wang's theorem (Ref. \cite{k4872},also see Ref. \cite{k0930}) states
that (for fixed $\lambda$) the set of $\pst$-invariant connections on
$P$ is in one-to-one correspondence with the set of linear maps
$\Lambda:\kf\ra\g$ that satisfy
\leqn{wang1}{
\begin{array}{ccll}
\Lambda(X)&=&\lambda
(X) & (X\in\kf_0),\\
\Lambda\circ ad_k &=& ad_{\lambda(k)}\circ\Lambda & (k\in K_0)
\end{array}
}
(where $\lambda$  now also denotes the induced Lie algebra homomorphism),
and the invariant connection and curvature on $P$ are then given by
\leqnarr{wang2ab}{
\ango\Xt,\omt\angc &\eqs& \Lambda(X)\phantom{[,\Lambda(Y)]-\Lambda([X,Y])} \quad (X\in\kf),\label{wang2a}\\
\ango\Xt\wedge\tilde{Y}/\Omt\angc &\eqs& [\Lambda(X),\Lambda(Y)]-\Lambda([X,Y])  \quad (X,Y\in\kf). \label{wang2b}
}
The symbol $\eqs$ indicates that these equations only hold at the fixed point $u_0\in P$.
The second equation of \eqref{wang1} becomes infinitesimally
\leqn{wangcond}{
\Lambda([X,Y]) = [\lambda(X),\Lambda(Y)] \quad \forall\; X\in\kf_0, \forall\; Y\in\kf.
}

For practical calculations we need to introduce a basis
$\{\eb_{\al}|\al=1\ldots m\}$ of the Lie algebra $\kf$ such that the
corresponding generators $\{\eb_a|a=1\ldots n\}$ span the tangent
space $T_{x_0}M$ at $x_0$ while the $\{\eb_\Gamma|\Gamma=n+1\ldots
m\}$ span the Lie subalgebra $\kf_0$. Thus, if the structure constants $c^{\la}_{\al\be}$ are introduced by
\leqn{strc}{ [\eb_{\al},\eb_{\be}] =
c^{\la}_{\al\be} \eb_{\la}, } then \leqn{strcsg}{ c^a_{\Gamma\Delta} =
0.
}
The infinitesimal generators $\ebb_a$ on $M$ corresponding to $\eb_a$
form a frame field in a neighborhood of $x_0$, but this will, in
general, only be global on $M$ if $M$ admits a simply transitive
isometry subgroup and is thus a group manifold. Let $\{\thb^a\}$ be
the local 1-form field dual to $\{\ebb_a\}$.

A pseudo-Riemannian metric $g$ on $M$ can now be written in the form
\leqn{met}{
g = g_{ab}\thb^a\otimes\thb^b,
}
where the components $\go_{ab} := g_{ab}(x_0)$ satisfy
\leqn{metcond}{
\go_{r(a} c^r_{b)\Gamma} = 0,
}
because of the $\ad_{K_0}$-invariance. The
coefficients of the Levi-Civita connection and the curvature tensor
with respect to this frame at $x_0$ are then given by
\leqnarr{cnxc/R}{
\Gamma^a_{bc} &\eqs& -\half c^a_{bc} + g^{ar}c^s_{r(b}g_{c)s} \label{cnxc}\\
R^a_{\phantom{a}bcd} &\eqs& \Gamma^a_{rc}\Gamma^r_{bd} - \Gamma^a_{rd}
\Gamma^r_{bc} - c^r_{cd}\Gamma^a_{br} + c^a_{b\Sigma}c^\Sigma_{cd}. \label{cnxR}
}
From here on the symbol $\eqs$ denotes equality at $x_0$ only.  Note
that neither the components of $g$, $\Gamma$, nor $R$ are constant on
$M$, in general.

Equation \eqref{cnxc} is easily derived from the equation
$\Lie_{\ebb_a}g=0$ stating that the $\ebb_a$ are infinitesimal
isometries, and that assumptions that the linear connection is metric
and symmetric. Equation \eqref{cnxR} is obtained most conveniently from Wang's
theorem applied to the bundle of pseudo-orthogonal frames over
$M$. Here, however, the principal bundle and the connection on it are
already fixed as well as the action of $K$ on the bundle, which is the
natural lift of the action on $M$. Thus \eqref{wang2a} fixes the Wang
map together with the requirement of zero torsion and \eqref{wang2b}
then leads to \eqref{cnxR} (cf. Ref. \cite{k0930}, Ch.~X). In a systematic
study of EYM-systems from a Kaluza-Klein perspective in Ref. \cite{k4286},
the Riemann tensor for metrics on homogeneous spaces is also
calculated in a very explicit form in terms of the structure constants
of the symmetry group by another method which leads to a different but
equivalent expression.

The gauge fields being invariant under a transitive symmetry group are
also determined by their values at just one point of $M$ which we take
to be the origin $x_0$. Their derivatives that occur in the
Yang-Mills equations can be computed using again Wang's theorem so
that the field equations are reduced to a purely algebraic form. Let
$\sigma$ be a local section of $P$, thus satisfying
$\pi\circ\sigma=\id_M$, and introduce the local gauge potential $A$
and the gauge field $F$ by
\leqn{AF}{ A = \sigma^*\omt, \qquad
F=\sigma^*\Omt.
}
Then we have the following lemma:

\begin{lem} Under the above
assumptions the Lie derivative of the gauge curvature $F$ at $x_0\in
M$ can be written in the form
\leqn{LieF}{ \Lie_{\Xb}F \eqs  [\Lambda(X),F] - [\ango\Xb,A\angc,F].
}
\end{lem}
\begin{proof}

Since $\Xb=\pi_*\circ\sigma_*\Xb=\pi_*\Xt$ the vector field
$\hat{X}=\sigma_*\Xb - \Xt$ is vertical on $P$. Now
$\Lie_{\Xb}F=\Lie_{\Xb}\sigma^*\Omt=\sigma^*\Lie_{\sigma_*\Xb}\Omt=\sigma^*(\Lie_{\Xt+\hat{X}}\Omt)=\sigma^*\Lie_{\hat{X}}\Omt$
in view of \eqref{ominv}. But
\leqn{LieOm}{ \Lie_{\hat{X}}\Omt =
\iota_{\hat{X}}d\Omt + d\iota_{\hat{X}}\Omt =
-\iota_{\hat{X}}[\omt\wedge\Omt] = -[\ango\hat{X},\omt\angc,\Omt] +
[\omt\wedge\iota_{\hat{X}}\Omt] =- [\ango\hat{X},\omt\angc,\Omt]
}
in view of the Bianchi identities, $d\Omt+[\omt\wedge\Omt]=0$, and the
fact that $\iota_Z\Omt=0$ for any vertical vector field $Z$. Pulling
back \eqref{LieOm} to $M$ by $\sigma$, $\Lie_{\Xb}F =
\sigma^*\Lie_{\hat{X}}\Omt=-\sigma^*[\ango\hat{X},\omt\angc,
\Omt]=-[\sigma^*\ango\hat{X},\omt\angc,F]$. But
$\sigma^*\ango\hat{X},\omt\angc=\sigma^*\iota_{\sigma_*{\Xb}}\omt -
\sigma^*\ango\Xt,\omt\angc =
\iota_{\Xb}\sigma^*\omt-\sigma^*\ango\Xt,\omt\angc \eqs \iota_{\Xb}A -
\Lambda(X)$ by \eqref{wang2a}.
\end{proof}

We choose now for the vector field $\Xb$ the local space-time frame vectors $\ebb_a$ and let
\leqn{F/A}{
A = A_b\thb^b, \qquad F = \half F_{ab}\thb^a\wedge\thb^b.
}
Then, introducing the (space-time) covariant derivatives
$F_{ab/c} =(\Lie_{\ebb_c}F)_{ab} + 2 F_{r[a}\Gamma^r_{b]c}$, together with
\eqref{LieF}, we have
\leqn{covF}{
F_{ab/c} \eqs [\Lambda_c-A_c,F_{ab}] + 2 F_{r[a}\Gamma^r_{b]c},
}
where $\Lambda_c := \Lambda(e_c)$.

Since the gauge-covariant derivative of $F$ is defined by
\leqn{gaugcovF}{
D_\alpha F_{\beta\gamma} =
\nabla_\alpha F_{\beta\gamma}+[A_\alpha,F_{\beta\gamma}],
}
we now find, interestingly, that the Yang-Mills equations,
$D^\lambda F_{\lambda\alpha}=0$ can be
written in these frame components without involving the gauge potentials,
\leqn{YMeq}{
[\Lambda^r,F_{ra}] + \Gamma^\ell_{ar}F_\ell^{\phantom{\ell}r}
+ F_{a\ell}\Gamma^\ell_{rs}g^{rs} \eqs 0.
}
In view of \eqref{wang2b}, the frame components $F_{ab}$ of the Yang-Mills
field are given by
\leqn{Fab}{
F_{ab} \eqs [\Lambda_a,\Lambda_b] - c^r_{ab}\Lambda_r
- c^\Sigma_{ab}\lambda_\Sigma.
}

Einstein's equations are also easily formulated in these frame components,
\leqn{EE}{
R_{ab} = \kappa T_{ab},
}
where $\kappa = 8\pi\;(\text{Newton's constant})$, the velocity of light is set
to unity,
\leqn{stren}{
T_{ab} = X_{ab}-\quart X^r_r g_{ab}, \quad X_{ab}:=
\ango F_{ar},F_b^{\phantom{b}r}\angc,
}
and $\ango,\angc$ represents a bi-invariant scalar product on the gauge
group Lie algebra $\g$. The stress energy tensor $T_{ab}$ has zero
trace, and the Ricci tensor components are obtained from \eqref{cnxR}.

All these equations hold only at the origin $x_0\in M$ and they form a
complicated algebraic system. For a given isometry group $K$ of
space-time and a chosen basis of $\kf$ the structure constants can be
considered fixed. The homomorphism $\lambda$ can be chosen arbitrarily
and then fixed. Possible choices are found by considering the
subgroups of the gauge group $G$ onto which there are homomorphisms
from the isotropy group $K_0$, in particular, imbeddings of $K_0$ in
$G$. This classification is discussed (for semisimple $K_0$ and
semisimple $G$) in Ref. \cite{k4779},Ref. \cite{k5221}. After the choice of a
particular homomorphism, equations \eqref{wang1} or, infinitesimally,
\eqref{wangcond}, i.e.
\leqn{Lambdacond}{
[\Lambda_a,\lambda_\Gamma] + c_{a\Gamma}^r\Lambda_r
= -c_{a\Gamma}^\Sigma\lambda_\Sigma,
}
must be solved for $\Lambda$ that is then substituted into \eqref{YMeq},
\eqref{Fab} and into Einstein's equations \eqref{EE}.

In the (most important) case of a reductive homogeneous space
$c_{a\Gamma}^\Sigma=0$ and \eqref{Lambdacond} is a
homogeneous linear system. Then $\Lambda$ can also be
regarded as an intertwining operator between two linear
representations of the isotropy group $K_0$ in the following way.
We have $\kf = \kf_0\oplus\mf$ as a vector space and the map
$\Lambda$ in \eqref{wang1} is fully determined by the linear map
$\bar{\Lambda}:\mf \ra \g$ that satisfies
\leqn{reduct}{
\bar{\Lambda}\circ\phi = \psi\circ\bar{\Lambda},
}
where $\phi:K_0\times\mf\ra\mf:(a,X)\mapsto \ad_a X$
and $\psi: K_0\times\g\ra\g:(a,Z)\mapsto\ad_{\lambda(a)}Z$. Then
$\bar{\Lambda}$ is an intertwining operator between these
representations of $K_0$, namely the adjoint representation $\phi$ on
$\mf$ and the representation $\psi$ on $\g$

Also, the $g_{ab}$ are arbitrary, subject to \eqref{metcond}. But not
all choices need lead to nonisometric space-times. One can reduce the
number of free parameters by bringing $g_{ab}$ into a canonical form
using basis transformations by automorphisms of $K$ that leave the
subgroup $K_0$ invariant.

\sect{EYMshst}{EYM equations in spatially homogeneous cosmological models}

Let $(M,g)$ now be an $n+1$-dimensional space-time manifold with an
isometry group $K$ whose orbits are $n$-dimensional space-like
hypersurfaces so that $M=\Sigma\times\RE$ with $K$ acting transitively
on $\Sigma$ and $K_0$ the isotropy subgroup at $x_0\in\Sigma$. We
choose to describe the metric by a coordinate time $t$ and a frame
field $\{\ebb_a\}$ of Killing vector fields on $\Sigma$,
\leqn{4met}{
g = -dt\otimes dt + g_{ab}\thb^a\otimes\thb^b.
}
Assume also that the
$\ebb_\Gamma$ ($\Gamma=n+1\ldots m$) vanish at a fixed point
$x_0\in\Sigma$.  It then follows that the $\Sigma_t$-coordinate
components of the frame vectors $\ebb_a$ do not depend on the time $t$,
so that
\leqn{et}{
[\partial_t,\ebb_{\al}] = 0 |\quad \forall\; \al=1\ldots m.
}

The connection and curvature components with respect to the local
space-time frame field $\{\ebb_0=\partial_t,\ebb_a\}$ can then be
calculated in the standard fashion. If
\leqn{Kab}{
K_{ab} = \half \dot{g}_{ab} } is the extrinsic curvature of the
hypersurfaces and a dot denotes the time derivative, we have for the
Ricci tensor components
\leqnarr{ric}{
R_{00} &\eqs& -g^{rs}\dot{K}_{rs} + K^r_s K^s_r, \\ R_{0b} &\eqs&
K_b^r c^s_{sr} + K^r_s c^s_{br}, \\ R_{ab} &\eqs& \dot{K}_{ab} + K^r_r
K_{ab} - 2 K_{ar}K^r_b + \RS_{ab}.
}
Here $\RS$ is the Ricci tensor
on $\Sigma$ and is given, according to \eqref{cnxR}, by
\leqn{ricn}{
\RS_{ab} \eqs \Gamma^r_{ab}\Gamma^s_{rs} - \Gamma^r_{sb}\Gamma^s_{ar}
- \Gamma^s_{ar}c^r_{sb} + c^r_{a\Sigma}c^\Sigma_{rb}.
}
 The $g_{ab}$ and $K_{ab}$ depend on $t$, the $c^{\al}_{\be\ga}$ are
constant (on \{$x_0\}\times\RE$) and the $\Gamma^a_{bc}$ are still
given by \eqref{cnxc}.

The calculation of the Yang-Mills equations for a gauge connection
invariant under a symmetry group with orbits on surfaces of constant
$t$ is analogous to the one on spherically symmetric static
space-times and is done as first outlined in Ref. \cite{k0832} (see also
Ref. \cite{hka26}). Locally one can introduce a gauge potential
$\mathcal{A}=A_0 dt+A$ where $A$ is the potential of a ($t$-dependent)
invariant connection on $\Sigma$ and $A_0$ is a $\g$-valued scalar,
invariant under $Ad_{\lambda(K_0)}$. In practice (unless there are
incompatible boundary conditions in the time evolution) $A_0$ can be
gauged away. This is because a time-dependent gauge transformation to
achieve such a result needs to satisfy an ordinary differential
equation on the gauge group that can always be solved, at least
locally in $t$.

In terms of the space-time co-frame $\{\thb^0=dt,\thb^a\}$ we now write for the Yang-Mills field
\leqn{FEB}{
F = E_a dt \wedge \thb^a + \half B_{ab}\thb^a\wedge\thb^b.
}
Then the Lie derivative of $F$ in the time direction is
\leqn{LieFt}{
\Lie_{\partial_t}F = \dot{E}_a dt\wedge\thb^a + \half \dot{B}_{ab}\thb^a\wedge\thb^b
}
and those along $\Sigma$ are still given by \eqref{LieF}. Just as in 
section~\ref{EYMhst} we can then compute the frame components of the covariant 
derivatives and find
\leqnarr{covFt}{
F_{ab/c} &\eqs& [\Lambda_c-A_c,B_{ab}] + 2 B_{r[a}\Gamma^r_{b]c} + 
2 E_{[a}K_{b]c}, \\
F_{0b/c} &\eqs& [\Lambda_c-A_c,E_b] - E_r\Gamma^r_{bc} + B_{br}K^r_c, \\
F_{ab/0} &\eqs& \dot{B}_{ab} + 2 B_{r[a}K^r_{b]}, \\
F_{0b/0} &\eqs& \dot{E}_b - E_r K^r_b.
}
The Yang-Mills equations thus become
\lgath{YMeqt}{
[E^r,\Lambda_r] - c^r_{rs} E^s \eqs 0, \label{YMeqt1}\\
\dot{E}_a + [A_0,E_a] + K^r_r E_a - 2K^r_aE_r - [B_{ar},\Lambda^r] + 
B_a^{\phantom{a}s}c^r_{rs} - \half g_{ar}c^r_{pq}B^{pq} \eqs 0, \label{YMeqt2}
}
where
\leqnarr{Bab/Ea}{
B_{ab} &\eqs& [\Lambda_a,\Lambda_b] - c^r_{ab}\Lambda_r - 
c^\Sigma_{ab}\lambda_\Sigma, \label{Bab} \\
E_a &\eqs& \partial_t\Lambda_a + [A_0,\Lambda_a] \label{Ea}
}
and we may choose the gauge such that $A_0=0$.

For the stress-energy tensor components we find (if we now restrict to $n=3$)
\leqnarr{strent}{
T_{00} &=& \half(E^2+B^2), \label{strent1} \\
T_{0a} &=& \epsilon_a^{\phantom{a}rs}\ango E_r,B_s\angc, \label{strent2} \\
T_{ab} &=& -\ango E_a,E_b\angc - \ango B_a,B_b\angc + \half(E^2+B^2)g_{ab}
 \label{strent3}
}
where $B_a:=\half \epsilon_a^{\phantom{a}rs}B_{rs}$, $E^2:=\ango
E_r,E^r\angc$, and $B^2:=\ango B_r,B^r\angc$. Einstein's equations
\eqref{EE} can now be brought into the form
\leqnarr{EEt}{
\RS + (K^r_r)^2 - K^{rs}K_{rs} &=& \kappa(E^2+B^2), \label{EEt1} \\
K_a^r c^s_{sr} + K^r_s c^s_{ar}  &=& \kappa\epsilon_a^{\phantom{a}rs}\ango E_r,B_s\angc, \label{EEt2} \\
\dot{K}_{ab} - 2 K_{ar}K^r_b + K^r_r K_{ab} + \RS_{ab} &=& \kappa T_{ab}. \label{EEt3}
}

If we choose the gauge such that $A_0=0$ then, after a basis of the
symmetry Lie algebra $\kf$ and the homomorphism $\lambda: K_0\ra G$
are chosen and a point $x_0\in\Sigma$ is fixed, we have as dynamical
variables the functions $g_{ab}(t)$, subject to \eqref{metcond}, and
the $\g$-valued functions $\Lambda_a(t)$, subject to
\eqref{Lambdacond}. Equations \eqref{EEt1} and \eqref{EEt2} can be
considered the Hamiltonian and the momentum constraints,
respectively. They restrict somewhat the choice of initial
values for an initial time but will afterwards be preserved by the
time evolution. This follows as a special case from the general analysis
of the Cauchy problem in EYM theory.

Only a time-independent basis transformation in $\kf$ by automorphisms
leaving $\kf_0$ invariant can now be used to possibly eliminate some
variables. The algebraic problem of finding the possible homomorphisms
$\lambda$ and solving for $\Lambda$ is similar to the one mentioned in
section~\ref{EYMhst} but a little simpler. The isotropy group $K_0$ is
now a subgroup of $SO(3)$ and thus compact so that the homogeneous
space is reductive. Moreover, on the three-dimensional space-like space
sections the isotropy group can only be either $SO(3)$ or $U(1)$ (or
trivial). We will consider in the following sections some of these
cases that can be handled without recourse to the more advanced techniques
of the theory of Lie algebra representations.

\sect{isotropic}{Isotropic cosmological models}

The isotropy subgroup $K_0$ of a space-time transitive isometry group
must be a subgroup of the Lorentz group and a classification of all
homomorphisms of such a subgroup into any compact gauge group $G$ is a
nontrivial algebraic problem. For a cosmological model with
three-dimensional homogeneous space sections the situation is much
simpler, since $K_0$ must be a subgroup of $SO(3)$ which leaves only
$SO(3)$, $U(1)$ or the trivial subgroup. In this section we consider
the ``physically isotropic'' models where $K_0$ is $SO(3)$. There are
still many possible conjugacy classes of homomorphisms $\lambda$ and
a complete classification for arbitrary compact
groups $G$ may not be known. We will here mainly consider the case when $G$ is
either $SU(n)$ or a real orthogonal group.

When $SO(3)$ is the isotropy group of an
isometric action on the three-dimensional manifold $\Sigma$ the
$(\Sigma,\gthree)$ must be of constant curvature $k$ and its isometry
group $K$ is $SO(4)$, $E(3)$ or $SO(3,1)$, respectively, depending on
whether $k$ is positive, zero or negative. The Lie
algebra has a basis $\{e_i,f_i\}\; (i=1\ldots3)$ with commutators
\leqnarr{constcurv}{
[e_i,e_j] &=& k\epsilon_{ij}^{\phantom{ij}r}f_r, \\{}%
[e_i,f_j] &=& \epsilon_{ij}^{\phantom{ij}r}e_r, \\{}%
[f_i,f_j] &=& \epsilon_{ij}^{\phantom{ij}r}f_r,
}
where the $f_i$ span the Lie algebra of the isotropy group. We can
choose $k$ to be $\pm1$ or $0$ and the $\epsilon_{ij}^{\phantom{ij}r}$ in this section now refers to the Euclidean metric in $\RE^3$.

The geometry of these isotropic models is then already determined,
namely the one of the well known Friedman-Robertson-Walker
space-times. We have in the terminology of section~\ref{EYMshst}
\leqn{frw}{
g_{ab} = a(t)\delta_{ab}, \quad K_{ab}=\half\dot{a}\delta_{ab},
\quad \RS_{ab}=2k\delta_{ab}
}
where the bar was dropped and $\{\theta^i\}$ is the co-frame dual to
$\{e_i\}$. In terms of the \emph{conformal time} $\tau$ the metric is
\leqn{conft}{
g = R(\tau)^2(-d\tau^2+\delta_{ab}\theta^a\otimes\theta^b)
}
so that $a=R^2$ and $\dot{\phi} = d\phi/dt = R^{-1}d\phi/d\tau=\dot{\phi}$ for
any function $\phi$. The stress tensor, being isotropic, is of the form
\leqn{Tiso}{
T_{ab}=p g_{ab}
}
 where $p$ is the pressure and, since the source will
be a zero-rest-mass Yang-Mills field, the mass-energy density is
$\mu=3p$. Einstein's equations are now equivalent to
\leqn{EEa}{
\ddot{a} = -2k \quad\text{and}\quad  \kappa p = \quart a^{-2}\dot{a}^2 + k a
}
or, in terms of the conformal time,
\leqn{EEconf}{
R'' + k R = 0 \quad\text{and}\quad  \kappa p = R^{-4}{R'}^2 + k R^{-2} = (\const)R^{-4}.
}
The complete time evolution of the geometry and thus the stress-energy
tensor is therefore easily obtained explicitly. It remains to formulate
the equations for the Yang-Mills field.

If we use again the notation $\Lambda_i = \Lambda(e_i)$ and
now $\lambda_i=\lambda(f_i)$ then equations \eqref{Lambdacond} become
\leqn{wangiso}{
[\lambda_i,\Lambda_j] = \epsilon_{ij}^{\phantom{ij}r}\Lambda_r.
}
They represent a system of linear equations for the $\Lambda_i$ once
the $\lambda_i$, i.e. the homomorphism is chosen. We have from \eqref{Bab} and \eqref{Ea}
\leqn{EBiso}{
E_i=\dot{\Lambda}_i = R^{-1}\Lambda'_i \quad\text{and}\quad  B_i = R^{-1}\left(
\half\epsilon_i^{\phantom{i}rs}[\Lambda_r,\Lambda_s] - k \lambda_i\right)
}
for the Yang-Mills fields (where the indices on $\Lambda$ and $\lambda$ are raised and
lowered with respect to $\delta_{ij}$) so that
\lgath{E2B2}{
E^2 = R^{-4}\delta^{rs}\ango\Lambda'_r,\Lambda'_s\angc \label{E2B21} \\
B^2 = \half R^{-4}\left( \ango[\Lambda_r,\Lambda_s],[\Lambda^r,\Lambda^s]\angc-4k\ango\Lambda_r,\Lambda^r\angc+2k^2\ango\lambda_r,\lambda^r\angc\right) \label{E2B22}
}
The YM field equations become
\lalign{YMiso}{
\Lambda_i'' - 2k \Lambda_i - [[\Lambda_i,\Lambda_r],\Lambda^r] = 0 \label{YMiso1} \\
[\Lambda'_r,\Lambda^r] = 0. \label{YMiso2}
}
From \eqref{EEt1}, \eqref{EEt2} and \eqref{Tiso} we have, moreover,
\lalign{EEiso}{
\epsilon_i^{\phantom{i}rs}\ango\lambda_r,\Lambda'_s\angc &= 0  \label{EEiso1}\\
\ango E_i,E_j\angc + \ango B_i,B_j\angc &= 2pg_{ij}. \label{EEiso2}
}
To derive these expressions we have used, whenever convenient, \eqref{wangiso} as well as the invariance of the inner product $\ango,\angc$ on $\g$.

We can go a little further before we need to specify the gauge group $G$, but the specific structure of the isotropy group and its action on $\Sigma$ incorporated in equations \eqref{wangiso} are essential. Equations \eqref{wangiso}
are a system of linear equations for the ($\g$-valued) $\Lambda_i$. Let $\{\La^K_i, K=0,\ldots,r-1\}$ be a basis of the solution space where $\La^0_i=\lambda_i$ since $\lambda_i$ is always a solution and is nonzero except if $\lambda$ is the trivial homomorphism.
\begin{lem}\label{lemIso}
The basis vectors $\{\La^K_i, K=0,\ldots,r-1\}$ of the solution space of Wang's conditions \eqref{wangiso} satisfy the following relations
\lalign{solsp}{
\epsilon_i^{\phantom{i}rs}[\La^K_r,\La^L_s] &= \gamma^{KL}_S\La^S_i \label{solsp1}\\
[\La^{(K}_i,\La^{L)}_j] &= \half\epsilon_{ij}^{\phantom{ij}r}\gamma^{KL}_S\La^S_r \label{solsp2}\\
\gamma^{KL}_M &= \gamma^{LK}_M \label{solsp3}\\
L^{KL} := \delta^{rs}[\La^K_r,\La^L_s] &= - L^{LK} \label{solsp4}\\
\ango\La^K_i,\La^L_j\angc &= \alpha^{KL}\delta_{ij} \quad\text{with}\quad\alpha^{KL}=\alpha^{LK} \label{solsp5}\\
\gamma^{KL}_S \alpha^{SM} &= \alpha^{KS}\gamma^{LM}_S \label{solsp6}\\
\gamma^{0K}_L = 2\delta^K_L & \quad\text{and}\quad  L^{0K} = 0 \label{solsp7}
}
\end{lem}
\begin{proof}
To prove \eqref{solsp1} let $L_i$ and $M_j$ be solutions of \eqref{wangiso} and $N_i=\epsilon_i^{\phantom{i}rs}[L_r,M_s]$. Then we can show that $[\lambda_i,N_j]=\epsilon_{ij}^{\phantom{ij}r}N_r$ by a simple calculation using the Jacobi identity and the identities satisfied by the Levi-Civita symbol $\epsilon_{ijk}$. Thus $N_i$ is also a solution of \eqref{wangiso}.
 (However, the full solution space need not be a Lie subalgebra of $\g$, in general.)
 \mnote{verified for $\su(2)\oplus\su(2)$!}\\
Equations \eqref{solsp2} and \eqref{solsp4} follow immediately from the antisymmetry of the Lie bracket and \eqref{solsp3} is a consequence of either \eqref{solsp1} or \eqref{solsp2}.\\
To prove \eqref{solsp5} we let $\alpha^{KL}_{ij}:=\ango\La^K_i,\La^L_j\angc$ and use \eqref{wangiso} and the invariance of the scalar product $\ango,\angc$,
$$\begin{array}{ccccccc}
\epsilon_{ij}^{\phantom{ij}r}\alpha^{KL}_{rk} &=& \ango\epsilon_{ij}^{\phantom{ij}r}\La^K_r,\La^L_k\angc &=& \ango[\lambda_i,\La^K_j],\La^L_k\angc
&=& -\ango\La^K_j,[\lambda_i,\La^L_k]\angc \\
&=& -\ango\La^K_j,\epsilon_{ik}^{\phantom{ik}r}\La^L_r\angc &=& -\epsilon_{ik}^{\phantom{ik}r}\alpha^{KL}_{jr},
\end{array}$$
from which the result easily follows.\\
Finally, \eqref{solsp6} follows directly from the invariance of the scalar product and \eqref{solsp7} is an immediate consequence of \eqref{wangiso} since $\Lambda^0_i=\lambda_i$.
\end{proof}

The only time-dependent quantities are now the amplitudes $\Phi_K(\tau)$ which satisfy the Yang-Mills equations in the form
\lgath{YMPhi}{
L^{KL}\Phi'_K\Phi_L = 0 \label{YMPhi1},\\
\Phi''_K - 2k\Phi_K + \half \gamma^{LM}_K\gamma^{PQ}_M\Phi_L\Phi_P\Phi_Q = 0 \label{YMPhi2}.
}
Here $(L^{KL})$, defined in \eqref{solsp4}, is an array of skewsymmetric matrices one for each dimension of the Lie algebra $\g$.
From \eqref{EBiso} we have
\leqn{nEBiso}{
E_i = R^{-1}\Phi'_K\La^K_i \quad\text{and}\quad  B_i = R^{-1}\left(\half\gamma^{KL}_M\Phi_K\Phi_L - k \delta^0_M\right)\La^M_i
}
and, in view of \eqref{solsp5}, Einstein's equations \eqref{EEiso1} and \eqref{EEiso2} reduce to \eqref{EEconf} and the following expression for the mass-energy density
\lgath{mener}{
\mu= \half(E^2+B^2)\label{mener1}\\
\intertext{where now}
E^2=3 R^{-4} \alpha^{KL}\Phi'_K\Phi'_L,\label{mener2}\\
B^2 = 3 R^{-4}\left(k^2\alpha^{00} - k \alpha^{0M}\gamma^{KL}_M\Phi_K\Phi_L + \quart\gamma^{KL}_R\alpha^{RS}\gamma^{PQ}_S\Phi_K\Phi_L\Phi_P\Phi_Q\right).\label{press3}
}
(Using the relations of Lemma~\ref{lemIso} it can be verified that $\mu R^4$ is constant as it should be.)

The quantities $\alpha^{KL}$, $\gamma^{KL}_M$ and $L^{KL}$ depend only
on the Lie algebra $\g$ and the homomorphism
$\lambda:\su(2)\ra\g$. Hence, to find all possible isotropic EYM
equations one has to find all $\su(2)$ subalgebras of $\g$ (up to
inner isomorphism), thus choosing the homomorphism $\lambda$ (see
Ref. \cite{k5221} and then solve the equation \eqref{reduct} for the
intertwining operator $\bar{\Lambda} = (\Lambda^K_i))$. This can be
done in a systematic way using a Cartan-Weyl basis of $\g$ by the
methods given in Ref. \cite{k4779}. Here we will only consider those
examples that can be dealt with in a more elementary way, without
involving the theory of Lie algebra root systems.

We know that all (connected) compact gauge groups
can be imbedded as subgroups of $GL(n,\RE)$ or $GL(n,\CO)$ (in fact in $SO(n))$
for some $n$. Moreover,
all finite-dimensional complex (real) representations of $SU(2)$ are
equivalent to unitary (real orthogonal) ones and decompose
orthogonally into irreducible parts. Thus at least for the unitary and
the real orthogonal groups we can determine the possible homomorphisms
directly from the well known representation theory.  If, for example,
$\tilde{\lambda}$ is a $n\times n$-unitary representation of $K_0$,
i.e. $\tilde{\lambda}:a\mapsto U_a\; \forall\; a\in K_0$ where $U_a$
is a unitary matrix, then $\lambda:a\mapsto (\det U_a)^{-1/n}U_a$ is a
homomorphism into $SU(n)$. Moreover, it is easily seen that equivalent
representations define conjugate homomorphisms and that, in fact,
\emph{conjugacy classes of homomorphisms of $K_0$ into $SU(n)$ are in
one-to-one correspondence with equivalence classes of $n$-dimensional
unitary representations of $K_0$}. Similarly, any real $n$-dimensional
orthogonal representation of $K_0$ immediately defines a homomorphism
into $SO(n)$.

If now $K_0=SU(2)$\cite{fnote1}
then any $n$-dimensional unitary (or real orthogonal) representation
is a direct sum of
irreducible unitary (real orthogonal) representations, i.e. any homomorphism
$\lambda:SU(2)\ra SU(n)$ is conjugate to one that maps into block
matrices
\leqn{block}{
\lambda(a) = \begin{pmatrix} D_{k_1}(a) & & \\ & \ddots & \\
& & D_{k_r}(a) \end{pmatrix}
}
where each $D_{k_i}$ is an irreducible $k_i$-dimensional
representation and where $k_1+\cdots+k_r=n$. As is well known, the Lie
algebra representation corresponding to an $n$-dimensional irreducible
representation can be written as follows. If
$\{\tau_1,\tau_2,\tau_3\}$ is the standard basis of $\su(2)$ in terms
of anti-Hermitian matrices and $\lambda_k=\lambda(\tau_k)$ are the images
in $\su(n)$ then the latter can be represented by the matrices
\leqnarr{ells}{
(\lambda_{+} )_{\ell m} &=& \sqrt{m(n-m)} \delta_{\ell,m+1},
\quad \lambda_{-} = \lambda_{+}^H \label{ells1} \\
\lambda_1 = -\ihalf(\lambda_{+}+\lambda_{-}), \quad
\lambda_2 &=& -\half(\lambda_{+}-\lambda_{-}), \quad
(\lambda_3)_{\ell m} = -i(\frac{n+1}{2}-m)\delta_{\ell m} \label{ells2}
}

Consider first a homomorphism class from $SU(2)$ to $SU(n)$, that
arises from an irreducible unitary representation in $\CO^n$. Then the
$\lambda_i$ in \eqref{wangiso} can be chosen as the matrices
\eqref{ells2} and the system \eqref{wangiso} can be explicitly
solved (this also follows from more general results of
representation theory) for the $\Lambda_j$ that can now be taken to
be $(n\times n)$ skew-Hermitian matrices. It follows that
\leqn{irred}{ \Lambda_i = \Phi \lambda_i }
i.e. the solution space is
one-dimensional. In this case the YM-potential is thus determined by a
single function $\Phi(\tau)$. For a simple Lie algebra like $\su(n)$
the invariant product $\ango,\angc$ must be a multiple of the Killing
form,
\leqn{kf}{ \ango X,Y\angc = -c_n\kappa(X,Y) }
for some constant
$c_n>0$ which we will choose to be $1$. It follows from
\eqref{solsp1} and \eqref{solsp2} that $\gamma^{00}_0=2$
and $\alpha^{00}=n^2(n^2-1)/6$ so that the Yang-Mills
equations become
\leqn{YMirrep}{ \Phi'' -2(k-\Phi^2)\Phi = 0 }
whence
\leqn{YMirrep2}{\frac{d\Phi}{\sqrt{c^2-(\Phi^2-k)^2}} = d\tau}
where the constant $c^2=4\mu R^4/(n^2(n^2-1))$. Thus $\Phi(\tau)$ is periodic in the cosmological time $\tau$ and can be expressed
in terms of an inverse elliptic integral. It is easily seen that the ``electric'' and ``magnetic'' contributions to the energy density $\mu$ oscillate in the time $\tau$. These equations (for $G=SU(2)$) have previously been derived and analyzed by Gal'tsov and Volkov \cite{k5209}.

If the homomorphism class is not induced by an irreducible
representation the gauge field may be more complicated. However, since
the evolution of the geometry of space-time is already determined only
the evolution of the gauge fields can be affected. Table~\ref{solsptab}
shows the dimensions $d$ of the solution space of
\eqref{wangiso} for the some homomorphisms $\lambda: \su(2)\ra\su(n)$. Here $1\oplus2$, for example, means that $\lambda$ is obtained from a representation in $\CO^3$ that decomposes into a (trivial) one-dimensional one and an irreducible two-dimensional one. In these
cases, according to \eqref{YMPhi2}, the YM field depends on $d$
independent amplitudes $\Phi_K(\tau)$ that each satisfy a second order equation. However, at least for $n\le6$, the $c$ constraint conditions
\eqref{YMPhi1}(which are not linearly independent in general) simply
imply that many of the $\Phi_K$ are proportional to each other so that
the remaining number $n_{\text{eq}}$ of second order equations that
must be solved is much smaller.

\begin{center}
\begin{table}[ht]
\caption{\footnotesize This table gives for different homomorphisms
$\lambda:\su(2)\ra\su(n)$
the number $d$ of dimensions of the solution space of \eqref{wangiso},
the number $c$ of nonzero constraint conditions \eqref{YMPhi1} and the
number $n_{\text{eq}}$ of independent amplitudes that satify second
order equations in time. (Trivial homomorphisms and those arising from
irreducible representations are not included.)} 
\label{solsptab}
\vspace{50mm}
\bc{\begin{tabular}{||c|c|c|c|c||}
\hline\hline
$n$ & $\lambda$ & $d$ & $c$ & $n_{\text{eq}}$ \\
\hline
3   & $1\oplus2$ & 1     & 0     & 1 \\
\hline
4   & $1\oplus1\oplus2$ & 1 & 0 & 1 \\
    & $1\oplus3$ & 3 & 1 & 2 \\
    & $2\oplus2$ & 4 & 3 & 2 \\
\hline
5   & $1\oplus1\oplus1\oplus2$ & 1 & 0 & 1 \\
    & $1\oplus1\oplus3$ & 5 & 6 & 2 \\
    & $1\oplus4$ & 1 & 0 & 1 \\
    & $1\oplus2\oplus2$ & 4 & 4 & 2 \\
    & $2\oplus3$ & 2 & 0 & 2 \\
\hline
6   & $1\oplus1\oplus1\oplus1\oplus2$ & 1 & 0 & 1\\
    & $1\oplus1\oplus1\oplus3$ & 7 & 11 & 2 \\
    & $1\oplus1\oplus4$ & 1 & 0 & 1 \\
    & $1\oplus1\oplus2\oplus2$ & 4 & 4 & 2 \\
    & $1\oplus2\oplus3$ & 4 & 3 & 3 \\
    & $1\oplus5$ & 1 & 0 & 1 \\
    & $2\oplus2\oplus2$ & 9 &11 & 2 \\
    & $2\oplus4$ & 4 & 4 & 3 \\
    & $3\oplus3$ & 4 & 5 & 2 \\
\hline\hline
\end{tabular}
}
\end{table}
\end{center}

To give one example, for an $SU(5)$-theory with the homomorphism $\lambda$
corresponding to a representation of the type $1\oplus1\oplus3$ we find the
Yang-Mills equations
\leqn{YM5}{
\begin{align}
\Phi'' + 2\Phi(\Phi^2+3\Psi^2-k) &= 0 \label{YM5_1} \\
\Psi'' + 2\Psi(3\Phi^2+\Psi^2-k) &= 0 \label{YM5_2} 
\end{align}
}
and
\leqn{YM5em}{
E^2 = 20R^{-4}\left({\Phi'}^2+{\Psi'}^2\right) \quad\text{and}\quad 
B^2 = 20R^{-4}\left[ \left(\Phi^2 + \Psi^2 - k \right)^2 + 4 \Phi^2\Psi^2\right].
}
The contribution of the electric and the magnetic part to the mass-energy density changes in time similarly as in the `irreducible' case, but the gauge fields now `rotate' in the Lie algebra in more dimensions.

If the gauge group is $SO(n)$ we can similarly classify the $\lambda$ by considering all n-dimensional real orthogonal representations of $\su(2)$. These decompose into irreducible blocks of dimensions $2k+1$ or $4k$ for integer $k$, but not $2k+2$ (see, e.g. Ref. \cite{k5930}).
It does not seem to be simple to write down formulae for these
representations for arbitrary $n$ as in \eqref{ells1} and
\eqref{ells2}. But there exists an algorithm to construct them
explicitly. First note that an irreducible complex representation of
$\su(2)$ leaves invariant a bilinear form $\beta$ on $\CO^n$. For the
choice of $\lambda$ in \eqref{ells1},\eqref{ells2} we find that
$\beta_{k\ell}=(-1)^k\delta_{\ell,n+1-k}$ which is symmetric for odd
$n$ and skew for even $n$.

Thus if $n$ is odd then $\lambda$ is of real type,
i.e. the representation is unitarily equivalent to one by real
orthogonal matrices. In fact,
\leqn{lareal}{
\tilde{\lambda}_k = U^H\lambda_k U \where U^HU=\id \quad\text{and}\quad 
 U^T\beta U=\id
}
are the generators of the orthogonal representation. The matrices $U$
can be easily computed by diagonalizing $\beta$ by congruence. For a
$\lambda:\su(2)\ra\so(2k+1)$ that corresponds to an irreducible
representation it now follows easily from the complex case that the
solutions of \eqref{wangiso} are again of the form \eqref{irred} and
the single time dependent amplitude $\Phi$ satisfies \eqref{YMirrep}.

For $n=4k$ the explicit irreducible representations are obtained via
the Lie algebra homomorphism
\leqn{co2re}{ \rho:\gl(\ell,\CO) \ra
\gl(2\ell,\RE):A=A_1+i A_2 \mapsto \tilde{A}=\begin{pmatrix} A_1 &
-A_2\\A_2 & A_1 \end{pmatrix}
}
which maps $\su(\ell)$ into $\so(2\ell)$. For $\ell=2k$ the image of
the matrices $\lambda_k$ generate an irreducible $4k$-dimensional real
orthogonal repesentation of $\su(2)$. Again, it can be verified
explicitly that \eqref{wangiso} has only the solutions \eqref{irred}
and that the only amplitude satisfies \eqref{YMirrep}.

The remaining equivalence classes of homomorphisms $\lambda$ into
$\so(n)$ can now be obtained from reducible orthogonal representations
in the same way as those for $\su(n)$. Some examples are tabulated in
Table~\ref{solsptabso}. The corresponding equations and expressions
for $E^2$ and $B^2$ are very similar to \eqref{YM5_1}, \eqref{YM5_2}, 
% \eqref{YM5} 
and \eqref{YM5em}.

\begin{center}
\begin{table}[ht]
\caption{\footnotesize Values $d$, $c$ and $n_{\text{eq}}$ for the 
equivalence classes of homomorphisms $\lambda:\su(2)\ra\so(n)$ for small $n$. 
Trivial homomorphisms and those arising from
irreducible representations are not included. The question marks indicate 
cases where the constraint equations do not simply imply that some amplitudes 
are proportional to others.} \label{solsptabso}

\vspace{50mm}
\bc{\begin{tabular}{||c|c|c|c|c||}
\hline\hline
$n$ & $\lambda$ & $d$ & $c$ & $n_{\text{eq}}$ \\
\hline
4   & $1\oplus3$ & 2 & 0 & 2 \\
\hline
5   & $1\oplus1\oplus3$ & 3 & 1 & 2 \\
    & $1\oplus4$ & 1 & 0 & 1 \\
\hline
6   & $1\oplus1\oplus1\oplus3$ & 4 & 3 & 2\\
    & $1\oplus1\oplus4$ & 1 & 0 & 1 \\
    & $1\oplus5$ & 1 & 0 & 1 \\
    & $3\oplus3$ & 3 & 1 & 2 \\
\hline
7   & $1\oplus1\oplus1\oplus1\oplus3$ & 5 & 6 & 2\\
    & $1\oplus1\oplus1\oplus4$        & 1 & 0 & 1\\
    & $1\oplus1\oplus5$               & 1 & 0 & 1\\
    & $1\oplus3\oplus3$               & 5 & 1 & ?\\
    & $3\oplus4$                      & 2 & 0 & 2\\
\hline
8   & $1\oplus1\oplus1\oplus1\oplus1\oplus3$ & 6 & 10 & 2 \\
    & $1\oplus1\oplus1\oplus1\oplus4$        & 1 & 0 & 1 \\
    & $1\oplus1\oplus1\oplus5$               & 1 & 0 & 1\\
    & $1\oplus1\oplus3\oplus3$               & 7 & 2 & ?\\
    & $1\oplus7$                             & 1 & 0 & 1\\
    & $3\oplus5$                             & 3 & 0 & 3\\
    & $4\oplus4$                             & 6 &17 & 2\\
\hline\hline
\end{tabular}
}
\end{table}
\end{center}

\sect{LRS}{Locally rotationally symmetric cosmological models}\label{lrs}

Spatially homogeneous cosmological models with $K_0=U(1)$ have been
extensively studied and are known as locally rotationally symmetric
(LRS) models.  Our construction of four-dimensional isometry groups of
LRS models is along the lines with Ref. \cite{k1263}.  It is known
that if $K_0$ is compact, then there exists a reductive
decomposition of $\mathfrak{k}$ (i.e. there is a subspace $\mathfrak{m}$ such 
that
$\mathfrak{k}=\mathfrak{k}_0+\mathfrak{m}$ and
$[\mathfrak{k_0},\mathfrak{m}]\subset\mathfrak{m}$ and
$\mathfrak{k}_0\cap\mathfrak{m}=0$). The choice of such a
reductive decomposition is not unique.  As it will be seen shortly, a
judicious choice of a reductive decomposition, greatly simplifies
the EYM equations.  It is interesting to note that for all Bianchi
cosmologies except Bianchi~III, there is a reductive decomposition in
which $\mathfrak{m}$ is a Lie subalgebra (such a decomposition for BIII would  
require $SU(1,1)$ to be solvable which contradicts the simplicity of 
$SU(1,1)$).  In a  
suitable basis
$e_1,\cdots,e_4$ such that $e_1,e_2,e_3$ span $\mathfrak{m}$ and
$e_4$ span $\mathfrak{k}_0$,
\leqn{commuteU(1)}{
-c^2_{14}=c^1_{24}=1,\;c^a_{34}=0 \qquad (a=1,2,3).
}
The Ad$(K_0)$-invariance of the metric expressed via (\ref{metcond}) then 
restricts the space-metric to the form $\mbox{diag}(f^2,f^2,f^2\sigma^2)$ 
where $f$ and $\sigma$ are 
functions of $t$. Given an invariant basis on a homogeneous space, one can 
start 
from this metric and, after integrating the Killing equations, find out which 
spatially homogeneous space-times admit the action of a four-dimensional 
isotropy 
group (cf. Table~\ref{table3} and Ref. \cite{k5508}).  Kramer {\it et al.} 
\cite{k1263} have 
classified all such 
space-times with two integers $\ell$ and $k$ (Bianchi~V (BV) does 
not 
fall into this category and is treated separately).  All
homogeneous spaces which have the same four-dimensional isometry group, belong 
to 
group
manifolds (Bianchi cosmologies).  Such group manifolds correspond to 
different
three-dimensional subgroups of the isometry group which act simply 
transitively on 
the
hypersurfaces of homogeneity.  Kramer {\it et al.'s} classification 
(section 11.1) 
is based on the metric
\leqn{metric in coordinates}{ 
\begin{array}{rcl}
g & = & f^2[ 2C^{-2}(dx^2+dy^2)+\quart \sigma^2 dz^2
        -\ell\sigma^2 C^{-1}(ydx-xdy)dz \\
  &   & \phantom{f^2[\ }+  \ell^2\sigma^2C^{-2} (ydx-xdy)^2], \\
  &   & \text{where\ \ } C := 1+1/2 k(x^2+y^2)\\[3pt]
\multicolumn{3}{l}{\text{or, for Bianchi V,}}\\[3pt]
g & = & f^2[e^{2z}(dy^2+dx^2)+\sigma^2 dz^2].
\end{array} }
These metrics all have (generically) four-dimensional isometry groups. 
We must now select a frame field of Killing vectors in such a way as to let 
$e_4$ generate the
isotropy group and the structure constants to satisfy \eqref{commuteU(1)}. 
The following choice achieves this.
\leqn{Kvfs in terms of l and k}{
\begin{array}{rclr}
e_1 & = & -\frac{k}{\sqrt{2}}xy\partial_y-\frac{1}{\sqrt{2}}(1+K)
\partial_x+\sqrt{2}\ell y\partial_z, & (\partial_x), \\[3pt]
e_2 & = & \phantom{-}\frac{k}{\sqrt{2}}xy\partial_x+\frac{1}{\sqrt{2}}(1-K)
\partial_y+\sqrt{2}\ell x\partial_z, & (\partial_y), \\[3pt]
e_3 & = & -2\partial_z, & (-x\partial_x-y\partial_y+\partial_z), \\[3pt]
e_4 & = & x\partial_y-y\partial_x, & (y\partial_x-x\partial_y),
\end{array} 
}
where $K:=(k/2)(x^2-y^2)$ and the entries of the right column are the Killing 
vector fields of BV.
The above Killing vector fields and non-vanishing structure constants
\leqn{structureconstants}{
c^3_{12}=\ell,\quad c^4_{12}=k \quad\mbox{or}\quad c^1_{13}=c^2_{23}=-1 
\mbox{ for BV},
}
determine the isometry group, embeddings of the isotropy group in the
isometry group up to conjugacy class, and identify the three-dimensional
homogeneous spaces that admit an action of a four-dimensional isometry group.
Here $\Sigma$  
is simply connected.  It is known that the number of degrees of freedom 
in mini-superspace models  
depends on the choice of topology \cite{k5844}.

\begin{center}
\begin{table}[ht]
\caption{\footnotesize The three-homogeneous cosmologies with a 
four-dimensional 
isometry group. $WH$ refers to Weyl-Heisenberg group.} 
\label{table3}
\vspace{50mm}
\bc{\begin{tabular}[h]{||l|l|l|c|c||}\hline\hline
Class & Homogeneous cosmology & Isometry group&$l$&$k$ \\ \hline
A     &BI              &$E(2) \otimes U(1)$ &$0$&$0$\\ \cline{1-2}
A     &$\mbox{BVII}_0$ &                   &&      \\ \hline
B     &BV     &$BVII_h\otimes U(1)$ &-&- \\ \cline{1-2}
B     &$\mbox{BVII}_h$       &                             &  & \\ \hline
B     &BIII              &$SU(1,1)\otimes U(1)$         &$0$&$-1$ \\ \cline{1-2}
A     &BVIII                 &                           &$1$&$-1$  \\ \hline
A     &BII    &$WH\otimes U(1)$         &$1$&$0$  \\ \hline
A     &BIX    &$SU(2)\otimes U(1)$ &$1$&$1$   \\ \hline
-     &Kantowski-Sachs       &$SU(2)\otimes R $    &$0$&1\\ \hline\hline
\end{tabular}
}
\end{table}
\end{center}

Our aim is to construct the invariant $SU(n)$-YM connections for homogeneous 
spaces listed
in the above table.  In doing so, we have to find all the conjugacy classes of 
homomorphisms
$\lambda:U(1)\rightarrow SU(n)$. 

Such conjugacy classes of homomorphisms are well understood 
for
\mnote{'static' is not necessary. The discussion in hka26 is also true for
time dependent systems}
spherically symmetric solutions of the EYM equations (cf. Ref. \cite{hka26}).  
These classes of
homomorphisms are basically of the same form as (\ref{block}).  However, 
since the
irreducible representations of $U(1)$ are one-dimensional, $D_k$ have only one
entry.  Therefore if $U(1)=\{z\in \CO:|z|=1\}$, then
\leqn{diag}{
\lambda:z\mapsto \mbox{diag}(z^{j_1},\cdots,z^{j_n})\quad 
(\sum_{i=1}^{n}j_i=0,\;
j_i=\mbox{an integer})
}
is clearly a homomorphism of $U(1)$ into $SU(n)$.  The set of integers $j_p\; 
(p=1,\cdots,n)$ such
that $j_p\geq j_q$ for $p<q$, yields all conjugacy classes of
homomorphisms $\lambda:U(1)\rightarrow SU(n)$.Denoting ${\cal D}:=
(i/2)\mbox{diag}(j_1,\cdots,j_n)$ we have
\leqn{wangU1}{
\Lambda[e_4,e_i]=[\lambda(e_4),\Lambda_i]=[{\cal D},\Lambda_i]
\quad\Longrightarrow\quad c^r_{4i}\Lambda_r=[{\cal D},\Lambda_i]
}
in which $\Lambda_i$ are traceless antihermitian matrices as in 
section~\ref{isotropic}.  These equations and (\ref{commuteU(1)}) give
\leqn{wang2}{
\Lambda_2=-[{\cal D},\Lambda_1],\quad \Lambda_1=[{\cal D},\Lambda_2],\quad 
[{\cal D},
\Lambda_3]=0,
}
which in turn yield
\leqn{wang3}{
(\Lambda_{l})_{pq}[4-(j_p-j_q)^2]=0,\quad l=(1,2).
}
The solution to the above equations is
\leqn{lamdaplus}{
\Lambda_1=i/2(\Lambda_+-\Lambda_-),\quad\Lambda_2=-1/2(\Lambda_++\Lambda_-),
\quad\Lambda_+=-(\Lambda_-)^H}
where $j_p\geq j_q$ for $p<q$ and therefore $\Lambda_+(\Lambda_-)$ is a
strictly upper (lower) triangular matrix.   Moreover, $(\Lambda_+)_{pq}\not=0$
only if  $j_p=j_q+2$.
The general solution of the above equations is in the root space corresponding
to
${\cal D}\subset$ (the Cartan subalgebra of $\su(n)$) and in principle could be 
obtained for any
compact group.  However, such a general treatment is out of the scope of the 
present
paper (cf. Ref. \cite{k5281}).
Some interesting special cases to consider are the following:
\begin{itemize}
\item[(a)]$j_p=0$, $\forall\;p\in \{1,\cdots,n\}$, (trivial homomorphism) 
requires
$\Lambda_1=\Lambda_2=0$ and $\Lambda_3$ is completely undetermined.
\item[(b)]If $|j_p-j_q|\not=2$ $\forall$ $p,q\in\{1,\ldots,n\}$ then
$\Lambda_1=\Lambda_2=0$ and $\Lambda_3$ is a diagonal traceless 
anti-Hermitian matrix. In
this case the gauge group reduces to its maximal torus (i.e. 
$U(1)\otimes\cdots
\otimes U(1)\subset SU(n)$).
\item[(c)]If $j_p=j_{p+1}+2$, $\forall\;p\in\{1,\cdots,n-1\}\Rightarrow
{\cal D}=(i/2)\diag(n-1,n-3,\cdots,-n+1)$.  Then (\ref{wang2}) and 
(\ref{wang3}) respectively  
imply that
$\Lambda_3$ is an anti-hermitian traceless diagonal matrix and
$(\Lambda_{+})_{p,p+1}=-(\Lambda^H_{-})_{p+1,p}$ are the only non-vanishing 
entries of
$\Lambda_\pm$.
\end{itemize}
In (b) the EYM equations for $SU(2)$-YM fields reduce to that of axially 
symmetric
electromagnetic fields and one can show that (a) and (b) are gauge 
equivalent \cite{hka29}.  We consider
(c) the simplest non-Abelian YM field in which the entries of ${\cal D}$ 
correspond to the
magnetic quantum numbers in the $n$-dimensional unitary representation of 
$SU(2)$.
Up to a gauge transformation, this representation yields the only possible
non-abelian connection for $SU(2)$-YM fields.
Therefore we derive the
EYM equations for this particular example starting with
\leqn{cmatrix}{
\begin{array}{rcl}
(\Lambda_+)_{p,p+1} & = &\omega_pe^{i\gamma_p},\quad p\in \{1,\cdots,n-1\} \\
\Lambda_3 & = & i\mbox{ diag }(\alpha_1,\ldots,\alpha_p-\alpha_{p-1},
\ldots,-\alpha_{n-1}).
\end{array}
} 
The YM constraints
(\ref{YMeqt1}) in terms of these variables are as follows
\leqn{YMconstraint}{
\omega_p^2\dot{\gamma}_p
+2\dot{\alpha}_p\sigma^{-2}\left\{\begin{array}{l} 0\\1\end{array} \right\}
=0.
}
Terms in the upper (lower) part of the braces refer to the `general' (BV) case. 
The YM dynamical  
equations (\ref{YMeqt2}) consist of
\leqn{YMdynamics}{
\begin{array}{l}
\ddot{\omega}_p+(f^{-1}\dot{f}+\sigma^{-1}\dot{\sigma})\dot{\omega}_p+
f^{-2}\omega_p\left(\sigma^{-2}\tilde{\alpha}^2_p+\half\tilde{W}_p-f^{2}
\dot{\gamma}^2_p-
\left\{
\begin{array}{l} k \\ \sigma^{-2} \end{array}
\right\}\right)=0, 
\\[5pt]
\ddot{\gamma}_p+(2\dot{\omega}_p\omega^{-1}_p+f^{-1}\dot{f}+\sigma^{-1}\dot
{\sigma})
\dot{\gamma}_p+
 2(f\sigma)^{-2}\tilde{\alpha}_p\left\{
\begin{array}{l} 0 \\ 1 \end{array}
\right\}=0, 
\\[5pt]
\ddot{\alpha}_p+(f^{-1}\dot{f}-\sigma^{-1}\dot{\sigma})\dot{\alpha}_p+
f^{-2}\tilde{\alpha}_p\omega^2_p-
\half\ell\sigma^2f^{-2}\left[W_p+p(n-p)k\right]\left\{
\begin{array}{l} 1 \\ 0
\end{array}
\right\}=0
\end{array}
}
and Einstein equations (\ref{EEt1}-\ref{EEt3}) are, respectively,
\leqn{EEASB}{\begin{array}{lcl}
3f^{-2}\dot{f}^2+2f^{-1}\dot{f}\sigma^{-1}\dot{\sigma}+f^{-2}\left
\{\begin{array}{l}
k-\quart\ell^2\sigma^2  \\
-3\sigma^{-2}
\end{array}\right\} & = & \kappa f^{-2}(T_1+T_2), 
\\[5pt]
\ddot{f}+2f^{-1}\dot{f}^2+\dot{f}\sigma^{-1}\dot{\sigma}+f^{-1}\left\{
\begin{array}{l}
k-\half\ell^2\sigma^2 \\
-2\sigma^{-2} \end{array} \right\} & = & \kappa f^{-1}T_1, 
\\[5pt]
\ddot{\sigma}+3f^{-1}\dot{f}\dot{\sigma}-f^{-2}(k\sigma-\ell^2\sigma^3)
\left\{\begin{array}{l}
1 \\
0 \end{array}\right\} & = & \kappa\sigma f^{-2}(T_2-2T_1).
\end{array} }
with the only non-trivial momentum constraint given by
\leqn{BVmomentum}{
\sigma^{-1}\dot{\sigma}\left\{\begin{array}{l} 0 \\ -1 \end{array} \right\}=
\kappa nf^{-2}\left(\sum_{p}\tilde{\alpha}_p\dot{\gamma}_p\omega^2_p-
\sum_{p}\omega_p\dot{\omega}_p\left\{
\begin{array}{l} 0 \\ 1 \end{array}\right\}\right).
}
Here we have used the abbreviations
\leqn{def1}{\begin{array}{lcl}
\tilde{\alpha}_p & := & 2\alpha_p-\alpha_{p-1}-\alpha_{p+1}, \\
W_p & := & \omega_p^2
-\left\{\begin{array}{l}
2\ell\alpha_p \\ 0
\end{array}\right\}, 
\\
\tilde{W}_p & := & 2W_p-W_{p-1}-W_{p+1}
+\left\{\begin{array}{l} 4k\\ 0 \end{array} \right\} .
\end{array}
}
and
\leqn{def2}{ \begin{array}{lcl}
T_1 & := & n\left[\sigma^{-2}{\DS\sum_p}\dot{\tilde{\alpha}}_p\dot{\alpha}_p +
\quart f^{-2}\left({\DS\sum_p}\tilde{W}_p W_p + 
(1/3)n(n^2-1)k^2\left\{ \begin{array}{l} 1 \\ 0 \end{array} \right\}
\right)
\right], 
\\
T_2 & := & n{\DS\sum_p}\left[\dot{\omega}_p^2 + \omega_p^2\dot{\gamma}^2_p +
\omega^2_p (f\sigma)^{-2}\left(\tilde{\alpha}^2_p +
\left\{ \begin{array}{l} 0 \\ 1 \end{array} \right\}
\right)\right],
\end{array}
}
and it is understood that all subscripted quantities are zero when the
index is outside the range $\{1,\ldots,n-1\}$.

At this point, we do not intend to give a complete analysis of the
above system of differential equations.  However, a few points are in
order. For the general case, if $\omega_p\neq0\;\forall p$,
$\dot{\gamma}_p=0$ and the first equation in (\ref{EEASB}), the
Hamiltonian constraint, is the only constraint of the system.  The
dynamical evolution is expected to preserve the constraint
$\dot{H}=0$.  Indeed, as a check on the consistency of the above
equations, one can show, for example for $G=SU(2)$, that
$\dot{H}=-(6\dot{f}/f+2\dot{\sigma}/\sigma)H$.
One observes that there are $2(n-1)$ degrees of freedom associated
with YM fields. Such an explicit integration is very complicated for
the Bianchi~V case, but as mentioned at the end of
section~\ref{EYMshst} we would expect the constraints to be conserved
in view of the general consistency of the Cauchy problem.

The above system is the set of $SU(n)$-EYM equations for the
particular homomorphism from $U(1)$ to $SU(n)$ chosen above for all
spatially homogeneous cosmologies with isotropy group $U(1)$.  These
equations are mildly gauge dependent ($A_0$ was set to $0$).  Nevertheless, the
gauge-invariant quantities like the various components of the
energy-momentum tensor, are easily expressible in terms of $\alpha_p$,
$\gamma_p$, and $\omega_p$.  We plan to pursue a more detailed
analysis of these equations for $SU(2)$-YM fields.

\sect{ac}{Acknowledgements}\label{ac1}
HPK would like to thank Othmar Brodbeck for useful discussions and 
BKD would like to thank Ali Mostafazadeh for the help in proof reading.

%\bibliography{mrabbrev,hkmrsupp,eym,proc}
\bibliography{mrabbrev}

\begin{thebibliography}{10}
\bibitem{fnote1}Perhaps one should take $SO(3)$ rather
than $SU(2)$ as the isotropy group. But it is clear that a covering
group of an isometry group of a Riemannian manifold $M$ will also act
in a well defined way via projections at least if $M$ is also simply
connected. We will not consider these topological questions here.
\end{thebibliography}
\end{document}